\documentclass{aa}  

\usepackage{graphicx}
\usepackage{txfonts}
\begin{document} 

\title{Intriguing detection of $^{12}$CO molecular emission\\in a classical Be star\thanks{Based on observations obtained 
1) at the Gemini Observatory, which is operated by the Association of Universities for Research in Astronomy, Inc., under a cooperative agreement with the NSF on behalf of the Gemini partnership: the National Science Foundation (United States), the National Research Council (Canada), CONICYT (Chile), the Australian Research Council (Australia), Minist\'erio da Ci\^encia, Tecnologia e Inova\c c\~ao (Brazil) and Ministerio de Ciencia, Tecnolog\'ia e Innovaci\'on Productiva (Argentina), under programs: GN-2010B-Q-02, GN-2016A-Q-96, GN-2017A-Q-84, and GN-2020B-Q-212, 
2) at the 6.5-m Magellan telescopes at Las Campanas Observatory, and 
3) at the Perek 2-m telescope at Ond\v{r}ejov Observatory}
}


   \author{Y.R. Cochetti\inst{1,2}
          \and
          M.L. Arias\inst{1,2}
          \and
          M. Kraus\inst{3}
          \and
          L.S. Cidale\inst{1,2}
          \and
          A.F. Torres\inst{1,2}
          \and
          A.Granada\inst{4}
          \and
          O.V. Maryeva\inst{3,5}
          }

   \institute{Departamento de Espectroscop\'ia, Facultad de Ciencias Astron\'omicas y Geof\'isicas, Universidad Nacional de La Plata\\ \email{cochetti@fcaglp.unlp.edu.ar}
   \and
        Instituto de Astrof\'isica de La Plata (CCT La Plata - CONICET, UNLP), Paseo del Bosque S/N, La Plata, B1900FWA, Buenos Aires, Argentina
        \and
        Astronomical Institute, Czech Academy of Sciences, Fri\v{c}ova 298, 251\,65 Ond\v{r}ejov, Czech Republic
        \and
        Laboratorio de Procesamiento de Se\~nales Aplicadas y Computaci\'on de Alto Rendimiento, Sede Andina, Universidad Nacional de R\'io Negro, Mitre 630, San Carlos de Bariloche, R8400AHN, R\'io Negro, Argentina
        \and 
        Sternberg  Astronomical Institute, Lomonosov Moscow State University, Universitetskij Pr. 13, Moscow 119992, Russia
        }

   \date{Received September 15, 1996; accepted March 16, 1997}

 
  \abstract
   {In the group of B stars with spectroscopic peculiarities, we can find the Be and the B[e] stars. The Be stars are early-type rapid rotators that present, as their principal characteristic, emission lines of hydrogen and singly ionized metals due to the presence of a gaseous envelope. The B[e] stars present in their spectra heterogeneous features that reveal the presence of regions with very different properties in a gaseous and dusty envelope.}
   {Our goal is to study the evolution of the disks around peculiar B stars through the variability of their physical properties and dynamical structure, as well as to set constraints on different models and disk forming mechanisms.}
   {Throughout the last decade, we have carried out temporal monitoring of a sample of objects in the near infrared using spectroscopic facilities at the Gemini and Las Campanas Observatories. In the present work, we focus on the classical Be star 12 Vul, for which also optical spectra have been collected quasi-simultaneously.}
   {We observed variability in the hydrogen line profiles of 12\,Vul, attributed to dissipating and building-up processes of the circumstellar envelope. Also, we found that this Be star presented the $^{12}$CO band heads in emission in one observation. The emission of this molecule has not been previously reported in a Be star, while it is a common feature among B[e] stars. We obtained parameters to describe the $^{12}$CO emitting region and propose different scenarios to explain this intriguing emission.}
   {}


   \maketitle
%

\section{Introduction}

The stars with a B spectral type form a diverse group, with stars in different subgroups that present spectroscopic peculiarities. One of these groups are the Be stars, which are early-type rapid rotators that present, as their principal characteristic, emission lines of hydrogen and singly ionized metals. The origin of these lines is attributed to the presence of a gaseous envelope. According to the observational evidence, the most accepted model for the envelope is a viscous decretion disk in Keplerian rotation \citep{Rivinius2013}.

Many of the Be stars present spectroscopic and photometric variability on different timescales: long-term variations that are linked to episodes of mass-ejection, leading to a decretion disk; short-term variations that could be a reflection of stellar pulsations or rotational modulation; and V/R cyclic variations (variation of the violet to red peak ratio in double-peaked emission line profiles) attributed to the precession of a global one-armed oscillation in the disk \citep{Okazaki1991}. In the near infrared (NIR) spectral region, the presence of the disk is revealed through a moderated flux excess and hydrogen recombination lines. Particularly, K-band spectra of Be stars display numerous lines of Pfund (Pf) series, together with lines of Brackett and Paschen series, where the main transitions are Br$\gamma$, Br$\delta$, and Pa$\alpha$. The profiles of these lines are sensitive to the physical properties and dynamical structure of their line-forming regions \citep{Marlborough1997,Cidale2000,Mennickent2009,Granada2010,CochettiPHD}. In this context, NIR spectroscopy of Be stars constitutes a valuable tool to study the origin and evolution of the disk.

There are other peculiar groups of objects with some similarities to Be stars. The B[e] stars present the typical emission spectra that characterize the Be stars, together with forbidden emission lines and a strong flux excess in the NIR and mid-IR spectral regions because of the presence of a circumstellar envelope of gas and dust. The B[e] star group includes the following objects in different evolutionary stages \citep{Lamers1998}: B[e] supergiants (B[e]SG), pre-main-sequence objects (Herbig B[e]), compact  planetary nebulae (cPNB[e]), symbiotic B[e] stars (SymB[e]), or unclassified B[e] type stars (unclB[e]). 
The heterogeneous spectral characteristics in this group of objects reveal the presence of regions with very different properties in the envelope: the region nearest to the star where the hydrogen emits, a region a little colder and further away from the central star where the metal lines and molecules would form, and an outermost cold region where the dust would be located \citep{Swings1973}. The dust becomes evident because of the strong IR excess, and the molecules could be observed mainly because of the emission of the CO molecule \citep{Kraus2009,Oksala2012,Liermann2014,Ilee2014}.

In this work, we report the first $^{12}$CO band emission ever seen in a classical Be star: 12 Vul. Parameters of the $^{12}$CO emitting region were obtained by modeling the spectrum. We discuss the evolution of the envelope of 12 Vul in the last years and propose different scenarios that could explain the presence of this $^{12}$CO emission. 

\section{12 Vul} \label{sec:12Vul}

12\,Vul (HD 187811) is a Be star of spectral type B2.5 V and a member of the Local Association or Pleiades moving group \citep{Eggen1975,Hoffleit1991}. It has been cataloged as a double-line spectroscopic binary by \citet{Chini2012}, with a 3.7 day period \citep{Eggen1975}. However, other authors have reported no evidence of the presence of a companion \citep{Wang2018,Horch2020}. Based on Hipparcos photometry, \citet{Hubert2000} reported that 12 Vul showed short-lived (100-200 days) outbursts, while \citet{Lefevre2009} described the photometric variability observed as unsolved. 

\citet{Lenorzer2002Diagram} proposed that the diagram $\log(\text{Hu}_{14}/\text{Br}\alpha)$ versus $\log(\text{Hu}_{14}/\text{Pf}\gamma)$, based on the intensity of hydrogen emission lines observed in the L-band, is a powerful tool to constrain the density and spatial distribution of circumstellar gas around hot stars. In this diagram, LBVs, B[e], and Be stars fall in different regions according to the optical depth of the circumstellar envelope. On the other hand, \citet{Mennickent2009} proposed a three-group classification for Be stars, based on the relative intensity of hydrogen emission lines observed in the NIR L-band. In plotting stars from different groups in the Lenorzer's diagram, they found that the groups' membership is related to the optical depth of the circumstellar envelope. Stars with similar emission intensities of their Br$\alpha$, Pf$\gamma$, and Humphreys lines belong to Group I (close to the optically thick envelopes in the Lenorzer's diagram), those with Br$\alpha$ and Pf$\gamma$ emissions more intense than Humphreys lines form Group II (distributed in a region of moderate or small optical depth), while Group III includes those where no line emission is detected (related to stars that have lost most of their envelopes).

The intensity of 12\,Vul's hydrogen emission lines observed in the NIR has been followed during the last decades. \citet{Mennickent2009} found that the intensity of Humphreys lines decreased from 1998 to 2003, and the star changed between Group I and II of their classification. At the same time, the position of 12\,Vul in the Lenorzer's diagram also changed. These results indicated that the optical depth of the envelope was changing between the different observations. 

The H$\alpha$ line profile is also variable. The Group II classification in 2003 coincides with a decrease in the H$\alpha$ line profile intensity. At that time, this line was transitioning from the double-peaked profile in emission observed in 2001, to the weak emission overimposed on the photospheric absorption observed in 2004. Between 2005 and 2009, the H$\alpha$ line profile remained in emission with a double-peaked line profile, with small intensity variations. After that, the intensity started to decrease, and in 2014 the line presented an absorption profile without emission features. This variability in the H$\alpha$ profile was reported by \citet{Sabogal2017}, who propose that 12\,Vul might have passed through dissipating and building-up processes of the circumstellar envelope. 

\begin{table}
\caption{NIR observing log for 12 Vul.}              
\label{table:obsIR}      
\centering                                      
\begin{tabular}{c c c}          
\hline\hline                        
Obs. date & Observatory & Spectrograph \\
\hline                                   
2010-09-15 & Gemini North & GNIRS-LS \\
2016-06-25 & Gemini North & GNIRS-XD \\
2017-06-04 & Las Campanas & FIRE     \\
2017-07-08 & Gemini North & GNIRS-XD \\
2020-09-06 & Gemini North & GNIRS-LS \\
\hline                                             
\end{tabular}
\end{table}

\begin{table}
\caption{Optical observing dates for 12 Vul.}              
\label{table:obsHa}      
\centering                                      
\begin{tabular}{l c}          
\hline\hline                        
Obs. date & Observatory \\
\hline                                   
2010-07-05 & Ond\v{r}ejov \\
2011-09-08 & BeSS$^*$ \\
2012-10-11 & Ond\v{r}ejov \\
2013-05-05 & BeSS$^*$ \\
2014-07-24 & BeSS$^*$ \\
2015-07-16 & Ond\v{r}ejov \\
2016-07-10 & BeSS$^*$ \\
2017-07-28 & BeSS$^*$ \\
2018-07-28 & BeSS$^*$ \\
2019-09-03 & BeSS$^*$ \\
2020-05-18 & Ond\v{r}ejov \\
\hline                                             
\end{tabular}
\tablefoot{For details on the instruments and observers see http://basebe.obspm.fr/basebe/.}
\end{table}

\section{Observations} \label{sec:obs}

Our work-group has been carrying out temporal monitoring of a sample of Be stars for more than a decade, with the aim of studying the origin and evolution of the disk as well as setting constraints on different models and disk forming mechanisms \citep{Granada2010,CochettiPHD}. As part of this monitoring, we have observed the star 12 Vul in the NIR spectral range. We present the observing dates in Table \ref{table:obsIR}.

We obtained NIR spectra from Gemini Observatory using GNIRS in 2010, 2016, 2017, and 2020. In 2010, we used the long-slit mode (LS) with the 32 l/mm disperser, a 0.3 arcsec slit, and the following cameras: a) the short blue camera to observe the H- and K-bands; and b) the short red camera to observe the L-band. In 2016 and 2017, we used a) the short blue camera with a 0.3 arcsec slit and the 32 l/mm disperser in cross dispersion mode (XD) to observe the J-, H-, and K-bands simultaneously and b) the long red camera with a 0.1 arcsec slit and the 10 l/mm disperser to observe the L-band. All of these configurations provide a spectral resolution of R\,$\sim$\,1800. In 2020, we used the 110.5 l/mm disperser and the long blue camera with a 0.10 arcsec slit to obtain a resolution of R\,$\sim$\,18000 in the range of 2.29 - 2.4 $\mu$m. During all observations, several ABBA sequences were taken, and a late-B or early-A type star was observed close enough in time and sky position to perform the telluric absorption correction. Flats and arcs were taken with every observation. The data reduction steps included the subtraction of the AB pairs, flat-fielding, telluric correction, and wavelength calibration. The reduction process was carried out with the {\sc IRAF}\footnote{IRAF is distributed by the National Optical Astronomy Observatory, which is operated by the Association of Universities for Research in Astronomy (AURA) under cooperative agreement with the National Science Foundation.} software package.

We also obtained another NIR spectrum from Las Campanas Observatory in 2017. The observation was performed using the Baade Telescope with the Folded port InfraRed Echellette (FIRE). The configuration used allowed us to obtain the spectrum in the range of $0.8-2.5\,\mu$m in one exposure, with R$\,\sim\,6000$. Together with the target spectrum, we obtained flats, arcs, and a spectrum of a standard star to correct for telluric absorption. The reduction was made using the pipeline\footnote{The pipeline software is written in IDL and builds on several external packages, including the SDSS idlutils, idlspec2d, Jason Prochaska's xidl package, and Mike Cushing's Spextool.} provided by the observatory, except for the redder order where the wavelength calibration was not good because of the small number of lines in the Th-Ar calibration lamp. In this order, we reduced the spectrum again using the IRAF software package to perform the wavelength calibration using the telluric lines. 

As complementary data, we obtained spectra in the optical range, around the H$\alpha$ line profile. Four spectra were obtained using the Coud\'{e} spectrograph \citep{2002PAICz..90....1S} attached to the Perek 2-m telescope at the Ond\v{r}ejov Observatory. Those from 2010 and 2012 were taken with the 830.77 l/mm grating and a SITe $2030\times 800$ CCD, and the others were taken with a PyLoN $2048\times 512$~BX CCD. With both CCDs, the spectra covered a range of about 6250--6750~\AA~with a resolution of $R~\simeq~13\,000$ in H$\alpha$. For wavelength calibration, a comparison spectrum of a ThAr lamp, was taken immediately after each exposure. Data reduction was performed using standard IRAF tasks. For the aim to show one spectrum per year covering the time interval for the NIR observations (2010-2020), we completed our data using spectra available in the BeSS database \citep{Neiner2011}. 

\begin{figure}
\centering
\includegraphics[width=\columnwidth]{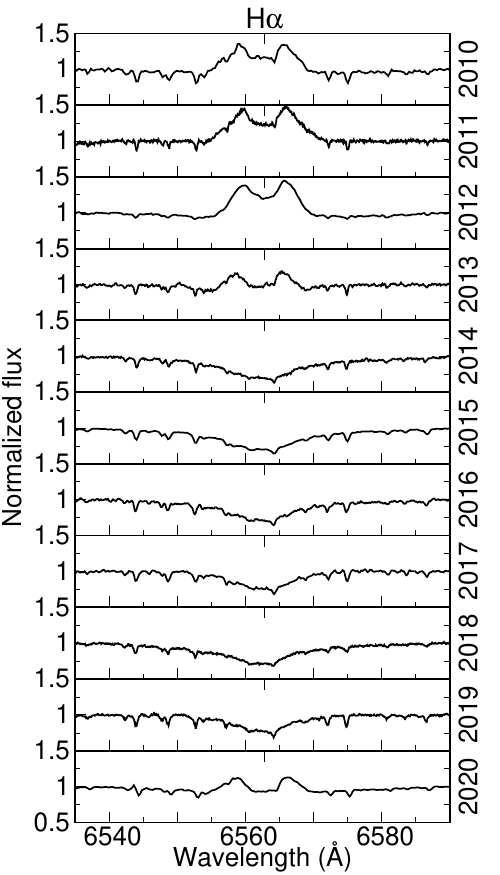}
\caption{Optical spectra of 12\,Vul around the H$\alpha$ line region. The H$\alpha$ line profile was in emission between 2010 and 2013. Between 2014 and 2019, there was no evidence of emission overimposed to the photospheric absorption. In 2020, the line profile was again in emission.}
\label{12Vul_Halpha}
\end{figure}

\begin{figure*}
\centering
\includegraphics[width=\textwidth]{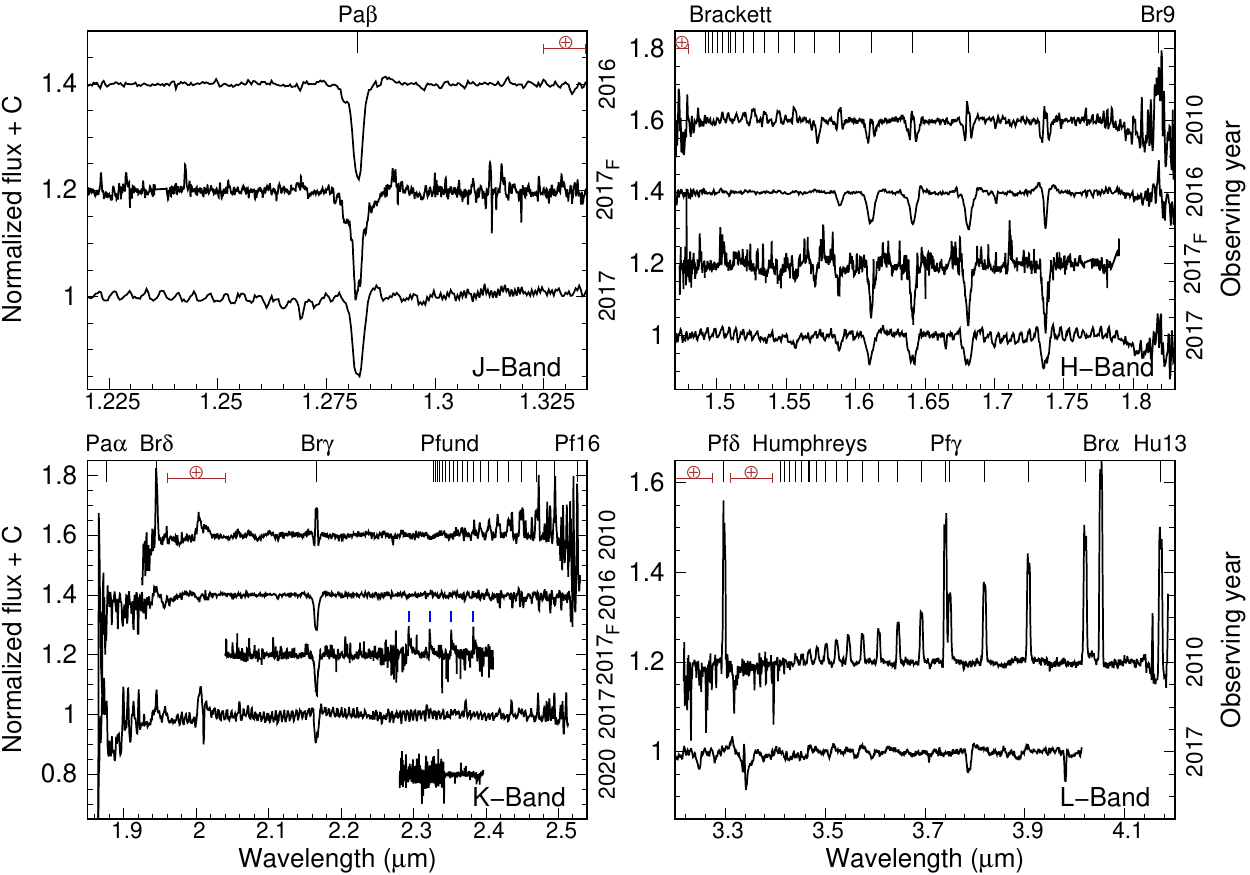}
\caption{NIR spectra of 12\,Vul in the J- (upper-left panel), H- (upper-right panel), K- (lower-left panel), and L-bands (lower-right panel). All the spectra were taken with GNIRS, except the one obtained with FIRE, which is pointed to with an F in the observing year. Spectra were normalized and vertically shifted for a better visualization. Positions of the hydrogen lines are marked in the top of each panel, and spectral ranges with telluric remnants are pointed out with the symbol $\oplus$. The $^{12}$CO band heads in emission in the K-band are indicated with blue tics.}
\label{12Vul_IR}
\end{figure*}

\begin{figure*}
\centering
\includegraphics[width=\textwidth]{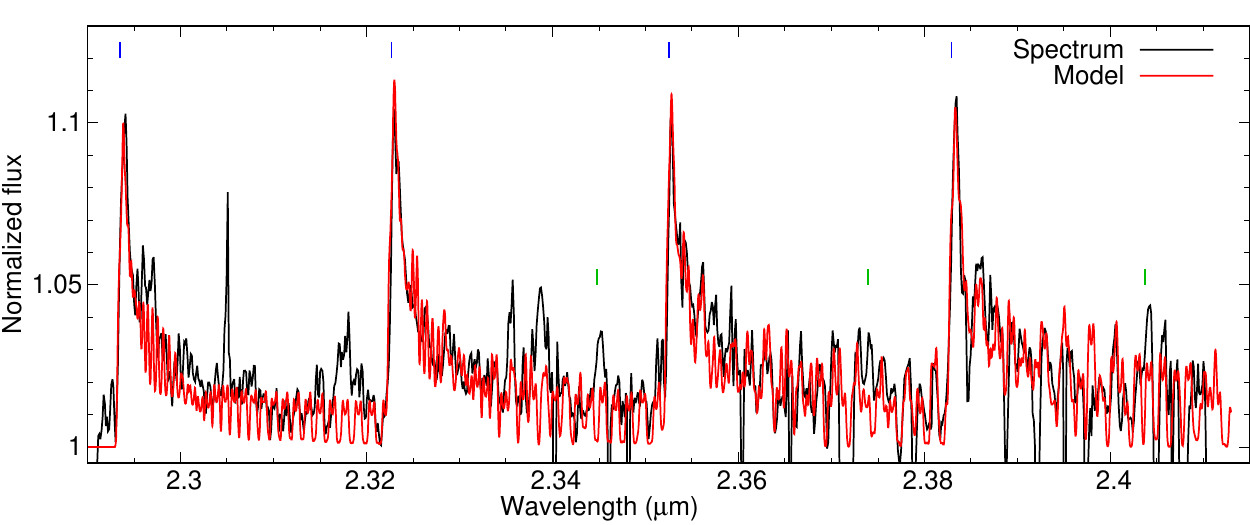}
\caption{Portion of the K-band spectrum of 12 Vul with $^{12}$CO emission (black). The synthetic spectrum is plotted in red. Blue and green tics point out the position of the $^{12}$CO and $^{13}$CO band heads, respectively.}
\label{fig:COmodel}
\end{figure*}

\section{Results} \label{sec:results}

\subsection{Hydrogen line profiles}

Figure~\ref{12Vul_Halpha} depicts one H$\alpha$ line profile of 12\,Vul per year between 2010 and 2020. From 2010 to 2013, H$\alpha$ was in emission. In the period from 2014 to 2019, the profile only showed photospheric absorption. The line profile was observed again in emission in our 2020 spectrum, but toward the end of the year the emission wings dissipated, according to BeSS spectra.

In Fig.~\ref{12Vul_IR} we show all NIR spectra of 12 Vul obtained in different epochs. In 2010, the line profiles of the Humphreys (L-band) and Pfund (K- and L-bands) series were observed in emission over the continuum level, while those of the Brackett series (H-band) were composite with a narrow emission overimposed on the broad photospheric absorption. In 2016 and 2017, hydrogen line profiles only showed weak emission over some photospheric profiles (see, for example, Pa$\beta$ line profile in the J-band).The spectrum taken in 2020 did not present evidence of hydrogen emission.

\subsection{CO molecular emission}

Surprisingly, the first 2017 spectrum (indicated as 2017$_{\text F}$) presented emission of $^{12}$CO. The $^{12}$CO band heads are pointed at with blue tics in Fig.~\ref{12Vul_IR}. No clear evidence of CO molecular emission was found in the spectrum taken one month later. However, the low resolution and bad signal-to-noise ratio of this spectrum could hinder CO detection.

The presence of $^{12}$CO molecular band emission in Be stars has not been previously reported. To reproduce the emission, we used the model proposed by \citet{Kraus2000}. This model assumes that the CO emission comes from a rotating ring or disk, with the CO in local thermodynamic equilibrium. The appearance of the band heads results from the superposition of ro-vibrational transitions, where each profile is double-peaked because of the rotational velocity. In addition to the rotational velocity, this model allows one to constrain the temperature and column density from the strength of the higher band head and the level of the quasi-continuum between them. 

Our fitting to the $^{12}$CO band spectrum is over-plotted to the stellar spectrum in Fig.~\ref{fig:COmodel}. This best fitting has been achieved for a narrow rotating ring of molecular gas with the following gas parameters: $V_{\text{rot}} (\text{projected to the line of sight}) = 42.5 \pm 2.5$~km~s$^{-1}$, $T_{\text{CO}} = 3250 \pm 250$ K, $N_{\text{CO}} = (7.5 \pm 2.5) \times 10^{20}$ cm$^{-2}$ (column density), and $V_{\text{turb}} = 1.5$~km~s$^{-1}$ (some intrinsic turbulent velocity on top of the thermal velocity of the CO gas). The synthetic spectrum was convolved to the spectral resolution.

There are a few peaks leftover in the spectrum, which could be real or artifacts. However, it is interesting to note that those at 2.3448, 2.3739, and 2.4037 $\mu$m coincide with the positions of $^{13}$CO. For modeling and to be sure that $^{13}$CO is present, we would need to obtain a better quality spectrum with CO emission.

\section{Discussion} \label{sec:dis}

\subsection{Evolution of the envelope}

According to the classification criteria from \citet{Mennickent2009}, 12\,Vul belonged to Group I in 2010, and to Group III in 2017. This means that since the observation in 2003 \citep{Mennickent2009}, the intensity of the Humphreys lines increased and in 2010, the star returned to Group I and changed  position in the Lenorzer's diagram again. This is in good agreement with the H$\alpha$ behavior since the Group I classification took place when this line had remained in emission for a few years. After that, the emission in all hydrogen features decreased and stayed undetectable for years until the 2020 H$\alpha$ observation. 

In the same way that \citet{Sabogal2017} attribute the variability of the H$\alpha$ emission and the position of 12\,Vul in the Lenorzer's diagram with dissipating and building-up processes of the circumstellar envelope, we could explain the variability in the profile of hydrogen lines observed in the last years with an episode of mass ejection before 2010 and with an envelope in a dissipation process in the following years. The weak emission in the H$\alpha$ line profile observed in 2020, which faded toward the end of the year, could provide evidence of stellar activity prior to a new disk building-up process.

\subsection{Proposed scenarios}

The sudden appearance of CO emission has been reported in the classical B[e] supergiant LHA 115-S 65 \citep{Oksala2012} and in some cataclismic variables \citep{Howell2008}. Though different in nature, both astrophysical objects hold a hot star with a circumstellar gaseous envelope (accretion or deccretion) that is cool enough at its outskirts to allow the formation or permanence of an extended dusty disk, which could be observed under particular conditions.

The presence of $^{12}$CO molecular band emission had not been previously reported in classical Be stars. For the density and temperature of the CO gas around 12 Vul, we found similar values as in various B[e] stars \citep{Kraus2016,Liermann2010,Kourniotis2018,Torres2018}. Moreover, the lack of CO gas with a temperature close to the CO dissociation value (5000 K) indicates that the molecular gas is detached from the central object \citep{Liermann2010}. While for the model calculation, we assumed that the CO gas is rotating, the observed double-peaked profiles of the individual CO lines might also be interpreted as an equatorial outflow with a constant velocity as in an expanding ring of ejected material.

We propose different scenarios to explain the $^{12}$CO molecular band emission observed in this Be star. One possibility is that the dissipating disk of 12\,Vul had reached either material previously expelled by the star during an outburst event, or even the interstellar, maternal CO cloud. The interaction, most likely accompanied by the compression and heating of the cool surrounding material, might provide suitable conditions in density and temperature. Alternatively, the formation of the disk structure around 12\,Vul could be caused by the interaction with an evolved cool companion and the binary system might evolve toward a configuration similar to that seen in some B[e] binaries. If true, 12\,Vul could be in a transition stage between a Be and a B[e] star.

To add evidence that could help to distinguish between the different possibilities, it would be interesting to follow up on this star with different techniques. For instance, radio observations could allow us to detect cold CO surrounding the star, which can be warmed by the dissipating disk, thus supporting the hypothesis of our first scenario. 

On the other hand, NIR observations would enable one to constrain the evolutionary stage of the star and support our second scenario. The $^{13}$C abundance on the stellar surface strongly increases during the main and post-main sequence evolution of massive stars, especially for rapidly rotating stars such as Be stars, and it changes the abundance in the wind due to the mass loss. A clear detection of $^{13}$C in the spectrum of 12 Vul would favor the hypothesis of a previously expelled material \citep{Kraus2009}.

Finally, finding evidence for the presence of a companion, for instance from speckle interferometry, it would help to characterize this object. This might be useful to understand its mass-loss phases and, therefore, the origin of the $^{12}$CO emission from this peculiar Be star.

\section{Conclusions}\label{sec:con}

We performed temporal spectral monitoring of the star 12 Vul between 2010 and 2020. In addition to the variability observed in the hydrogen line profiles, attributed to dissipating and building-up processes of the circumstellar envelope, we found CO band head emission in the K-band. We highlight that this surprising detection is reported for a classical Be star for the first time.

A reasonable fitting to the observed $^{12}$CO band emission was achieved with a narrow ring of gas either revolving around or expanding from the star with a velocity, projected to the line of sight, of 42.5$\pm$ 2.5\,km\,s$^{-1}$. Such a ring of molecular gas might result from the interaction of the dissipating disk with material from either a previous ejection or from the natal molecular cloud. Alternative scenarios related to close binary interactions also need to be explored to understand the origin of CO emissions and to discover possible links with the B[e] stars.

The discovery of hot CO gas in a classical Be star motivates us to undertake intensive spectroscopic follow up observations of 12 Vul in different wavelengths to gain a deeper insight into the envelopes of this object. In addition, it would be interesting to explore if this phenomenon is present in other Be stars. 

\begin{acknowledgements}
We thank the anonymous referee for the valuable comments that helped to improve our manuscript.

This work has made use of the BeSS database, operated at LESIA, Observatoire de Meudon, France: http://basebe.obspm.fr

We thank Lenka Kotkov\'{a} and Petr \v{S}koda for taking the H$\alpha$ spectra in Ond\v{r}ejov in 2010, 2012, and 2015. 

Y.R.C. thank Gabriel Ferrero and Nidia Morrell for allowing her to obtain data in Las Campanas Observatory during the nice stay shared, and acknowledges the support from the Carnegie Institution for Science and Richard Lounsbery Foundation that enable the stay in La Serena and Las Campanas Observatory under the visiting fellowship program for young Argentinian astronomers.

M.L.A. and A.F.T. acknowledges financial support from the Universidad Nacional de La Plata (Programa de Incentivos 11/G160).

L.S.C. thanks financial support from CONICET (PIP 0177) and from the Agencia Nacional de Promoci\'on Cient\'ifica y Tecnol\'ogica de Argentina (Pr\'estamo BID PICT 2016-1971)

A.G. acknowledges financial support from the Agencia Nacional de Promoci\'on Cient\'ifica y Tecnol\'ogica de Argentina (PICT 2017-3790).

M.K. acknowledges financial support from the Grant Agency of the Czech Republic (GA \v{C}R, grant number 20-00150S).

O.M. acknowledges support from the Czech Science Foundation GA18-05665S.

The Astronomical Institute Ond\v{r}ejov is supported by the project RVO:67985815. This project has received funding from the European Union's Framework Programme for Research and Innovation Horizon 2020 (2014-2020) under the Marie Sk\l{}odowska-Curie Grant Agreement No. 823734.
\end{acknowledgements}

   \bibliographystyle{aa} 
   \bibliography{biblio.bib} 

\begin{thebibliography}{30}
\expandafter\ifx\csname natexlab\endcsname\relax\def\natexlab#1{#1}\fi

\bibitem[{{Chini} {et~al.}(2012){Chini}, {Hoffmeister}, {Nasseri}, {Stahl}, \&
  {Zinnecker}}]{Chini2012}
{Chini}, R., {Hoffmeister}, V.~H., {Nasseri}, A., {Stahl}, O., \& {Zinnecker},
  H. 2012, \mnras, 424, 1925

\bibitem[{{Cidale} {et~al.}(2000){Cidale}, {Zorec}, {Maillard}, \&
  {Morrell}}]{Cidale2000}
{Cidale}, L., {Zorec}, J., {Maillard}, J.~P., \& {Morrell}, N. 2000, in
  Astronomical Society of the Pacific Conference Series, Vol. 214, IAU Colloq.
  175: The Be Phenomenon in Early-Type Stars, ed. M.~A. {Smith}, H.~F.
  {Henrichs}, \& J.~{Fabregat}, 472

\bibitem[{{Cochetti}(2019)}]{CochettiPHD}
{Cochetti}, Y.~R. 2019, PhD thesis, Facultad de Ciencias Astron\'omicas y
  Geof\'isicas, Universidad Nacional de La Plata

\bibitem[{{Eggen}(1975)}]{Eggen1975}
{Eggen}, O.~J. 1975, \pasp, 87, 37

\bibitem[{{Granada} {et~al.}(2010){Granada}, {Arias}, \&
  {Cidale}}]{Granada2010}
{Granada}, A., {Arias}, M.~L., \& {Cidale}, L.~S. 2010, \aj, 139, 1983

\bibitem[{{Hoffleit} \& {Jaschek}(1991)}]{Hoffleit1991}
{Hoffleit}, D. \& {Jaschek}, C. 1991, {The Bright star catalogue}

\bibitem[{{Horch} {et~al.}(2020){Horch}, {van Belle}, {Davidson}, {Willmarth},
  {Fekel}, {Muterspaugh}, {Casetti-Dinescu}, {Hahne}, {Granucci}, {Clark},
  {Winters}, {Rupert}, {Weiss}, {Colton}, {Nusdeo}, \& {Henry}}]{Horch2020}
{Horch}, E.~P., {van Belle}, G.~T., {Davidson}, James~W., J., {et~al.} 2020,
  \aj, 159, 233

\bibitem[{{Howell} {et~al.}(2008){Howell}, {Hoard}, {Brinkworth}, {Kafka},
  {Walentosky}, {Walter}, \& {Rector}}]{Howell2008}
{Howell}, S.~B., {Hoard}, D.~W., {Brinkworth}, C., {et~al.} 2008, \apj, 685,
  418

\bibitem[{{Hubert} {et~al.}(2000){Hubert}, {Floquet}, \& {Zorec}}]{Hubert2000}
{Hubert}, A.~M., {Floquet}, M., \& {Zorec}, J. 2000, in Astronomical Society of
  the Pacific Conference Series, Vol. 214, IAU Colloq. 175: The Be Phenomenon
  in Early-Type Stars, ed. M.~A. {Smith}, H.~F. {Henrichs}, \& J.~{Fabregat},
  348

\bibitem[{{Ilee} {et~al.}(2014){Ilee}, {Fairlamb}, {Oudmaijer},
  {Mendigut{\'\i}a}, {van den Ancker}, {Kraus}, \& {Wheelwright}}]{Ilee2014}
{Ilee}, J.~D., {Fairlamb}, J., {Oudmaijer}, R.~D., {et~al.} 2014, \mnras, 445,
  3723

\bibitem[{{Kourniotis} {et~al.}(2018){Kourniotis}, {Kraus}, {Arias}, {Cidale},
  \& {Torres}}]{Kourniotis2018}
{Kourniotis}, M., {Kraus}, M., {Arias}, M.~L., {Cidale}, L., \& {Torres}, A.~F.
  2018, \mnras, 480, 3706

\bibitem[{{Kraus}(2009)}]{Kraus2009}
{Kraus}, M. 2009, \aap, 494, 253

\bibitem[{{Kraus} {et~al.}(2016){Kraus}, {Cidale}, {Arias}, {Maravelias},
  {Nickeler}, {Torres}, {Borges Fernandes}, {Aret}, {Cur{\'e}},
  {Vallverd{\'u}}, \& {Barb{\'a}}}]{Kraus2016}
{Kraus}, M., {Cidale}, L.~S., {Arias}, M.~L., {et~al.} 2016, \aap, 593, A112

\bibitem[{{Kraus} {et~al.}(2000){Kraus}, {Kr{\"u}gel}, {Thum}, \&
  {Geballe}}]{Kraus2000}
{Kraus}, M., {Kr{\"u}gel}, E., {Thum}, C., \& {Geballe}, T.~R. 2000, \aap, 362,
  158

\bibitem[{{Lamers} {et~al.}(1998){Lamers}, {Zickgraf}, {de Winter}, {Houziaux},
  \& {Zorec}}]{Lamers1998}
{Lamers}, H. J.~G.~L.~M., {Zickgraf}, F.-J., {de Winter}, D., {Houziaux}, L.,
  \& {Zorec}, J. 1998, \aap, 340, 117

\bibitem[{{Lef{\`e}vre} {et~al.}(2009){Lef{\`e}vre}, {Marchenko}, {Moffat}, \&
  {Acker}}]{Lefevre2009}
{Lef{\`e}vre}, L., {Marchenko}, S.~V., {Moffat}, A.~F.~J., \& {Acker}, A. 2009,
  \aap, 507, 1141

\bibitem[{{Lenorzer} {et~al.}(2002){Lenorzer}, {de Koter}, \&
  {Waters}}]{Lenorzer2002Diagram}
{Lenorzer}, A., {de Koter}, A., \& {Waters}, L.~B.~F.~M. 2002, \aap, 386, L5

\bibitem[{{Liermann} {et~al.}(2010){Liermann}, {Kraus}, {Schnurr}, \& {Fernand
  es}}]{Liermann2010}
{Liermann}, A., {Kraus}, M., {Schnurr}, O., \& {Fernand es}, M.~B. 2010,
  \mnras, 408, L6

\bibitem[{{Liermann} {et~al.}(2014){Liermann}, {Schnurr}, {Kraus}, {Kreplin},
  {Arias}, \& {Cidale}}]{Liermann2014}
{Liermann}, A., {Schnurr}, O., {Kraus}, M., {et~al.} 2014, \mnras, 443, 947

\bibitem[{{Marlborough} {et~al.}(1997){Marlborough}, {Zijlstra}, \&
  {Waters}}]{Marlborough1997}
{Marlborough}, J.~M., {Zijlstra}, J.~W., \& {Waters}, L.~B.~F.~M. 1997, \aap,
  321, 867

\bibitem[{{Mennickent} {et~al.}(2009){Mennickent}, {Sabogal}, {Granada}, \&
  {Cidale}}]{Mennickent2009}
{Mennickent}, R.~E., {Sabogal}, B., {Granada}, A., \& {Cidale}, L. 2009, \pasp,
  121, 125

\bibitem[{{Neiner} {et~al.}(2011){Neiner}, {de Batz}, {Cochard}, {Floquet},
  {Mekkas}, \& {Desnoux}}]{Neiner2011}
{Neiner}, C., {de Batz}, B., {Cochard}, F., {et~al.} 2011, \aj, 142, 149

\bibitem[{{Okazaki}(1991)}]{Okazaki1991}
{Okazaki}, A.~T. 1991, \pasj, 43, 75

\bibitem[{{Oksala} {et~al.}(2012){Oksala}, {Kraus}, {Arias}, {Borges Fernand
  es}, {Cidale}, {Muratore}, \& {Cur{\'e}}}]{Oksala2012}
{Oksala}, M.~E., {Kraus}, M., {Arias}, M.~L., {et~al.} 2012, \mnras, 426, L56

\bibitem[{{Rivinius} {et~al.}(2013){Rivinius}, {Carciofi}, \&
  {Martayan}}]{Rivinius2013}
{Rivinius}, T., {Carciofi}, A.~C., \& {Martayan}, C. 2013, \aapr, 21, 69

\bibitem[{{Sabogal} {et~al.}(2017){Sabogal}, {Ubaque}, {Garc{\'{\i}}a-Varela},
  {{\'A}lvarez}, \& {Salas}}]{Sabogal2017}
{Sabogal}, B.~E., {Ubaque}, K.~Y., {Garc{\'{\i}}a-Varela}, A., {{\'A}lvarez},
  M., \& {Salas}, L. 2017, \pasp, 129, 014203

\bibitem[{{\v{S}lechta} \& {\v{S}koda}(2002)}]{2002PAICz..90....1S}
{\v{S}lechta}, M. \& {\v{S}koda}, P. 2002, Publications of the Astronomical
  Institute of the Czechoslovak Academy of Sciences, 90, 1

\bibitem[{{Swings}(1973)}]{Swings1973}
{Swings}, J.~P. 1973, \aap, 26, 443

\bibitem[{{Torres} {et~al.}(2018){Torres}, {Cidale}, {Kraus}, {Arias},
  {Barb{\'a}}, {Maravelias}, \& {Borges Fernandes}}]{Torres2018}
{Torres}, A.~F., {Cidale}, L.~S., {Kraus}, M., {et~al.} 2018, \aap, 612, A113

\bibitem[{{Wang} {et~al.}(2018){Wang}, {Gies}, \& {Peters}}]{Wang2018}
{Wang}, L., {Gies}, D.~R., \& {Peters}, G.~J. 2018, \apj, 853, 156

\end{thebibliography}

\end{document}